\documentclass[twocolumn,3p,times,procedia]{elsarticle}

\usepackage{ecrc}
\usepackage{graphicx}
\usepackage{titlesec}
\usepackage{multirow} %multirow for format of table 
\usepackage{xcolor}
\usepackage{algorithmic}
\usepackage[linesnumbered,boxed,ruled,commentsnumbered]{algorithm2e}
\usepackage{float}
\usepackage{ulem}

\volume{00}

\firstpage{1}

\runauth{}

\jid{procs}

\CopyrightLine{2021}{Published by Elsevier Ltd.}

\usepackage{amssymb}
\usepackage{amsmath}
\biboptions{sort&compress}

\usepackage[figuresright]{rotating}
\usepackage{bm}
\usepackage{caption}

\usepackage{fancyhdr}
\fancyheadoffset[RO,EL]{0pt}
\usepackage{geometry}
\usepackage[numbers,sort&compress]{natbib}

\titlespacing{\section}{0pt}{1.8ex plus .9ex minus .6ex}{.3ex plus .0ex}
\titlespacing{\subsection}{0pt}{1.5ex plus 1ex minus .1ex}{.7ex plus .2ex}

\begin{document}

% \listoffigures
\captionsetup[figure]{labelfont={bf},labelformat={default},labelsep=period,name={Figure}}
\begin{frontmatter}




\dochead{}
%% e.g. \dochead{17th International Conference on Dynamical Processes in Excited States of Solids}
\title{
\begin{flushleft}
{\bf  A Credibility-aware Swarm-Federated Deep Learning Framework in Internet of Vehicles}
\end{flushleft}
}

% \author[]{\bf \Large \leftline {Zhe Wang$^*$$^a$, Xinhang Li$^a$, Chen Xu$^a$, Lin Zhang$^a$}$^b$}

% \address{\bf  \leftline {$^a$Beijing University of Posts and Telecommunications, }
% \bf  \leftline{Beijing 100876, China}
% }
\author[]{\bf \Large \leftline {Zhe Wang$^a$, Xinhang Li$^a$, Tianhao Wu$^*$$^a$, Chen Xu$^a$, Lin Zhang$^{ba}$}}

\address{\bf  \leftline {$^a$School of Artificial Intellegence, Beijing University of Posts and Telecommunications, Beijing 100876, China}
% \bf  \leftline{Beijing 100876, China}

\bf  \leftline {$^b$Beijing Information Science and Technology University, Beijing 100096, China}

}

% \bf  \leftline {$^b$College of Computer Science and Engineering,
% Chongqing University of Technology, Chongqing 400054, China}

% \bf  \leftline {$^c$School of Computer Science and Engineering,
% Beihang University, Beijing, 100191, China}

\cortext[]{Corresponding Author. \emph{Email:} shogun2015@hotmail.com}

% \cortext[]{Corresponding Author. \emph{Email:} wu.tianhao@bupt.edu.cn}

% \fntext[]{a simple introduction about Si Li (email:).}
% \sout{}
% \textcolor{red}{\sout{}}
% \fntext[]{a simple introduction about San Zhang (email:).}
\begin{abstract}
Federated Deep Learning (FDL) is helping to realize distributed machine learning in the Internet of Vehicles (IoV). However, FDL’s global model needs multiple clients to upload learning model parameters, thus still existing unavoidable communication overhead and data privacy risks. The recently proposed Swarm Learning (SL) provides a decentralized machine-learning approach uniting edge computing and blockchain-based coordination without the need for a central coordinator. This paper proposes a Swarm-Federated Deep Learning framework in the IoV system (IoV-SFDL) that integrates SL into the FDL framework. The IoV-SFDL organizes vehicles to generate local SL models with adjacent vehicles based on the blockchain empowered SL, then aggregates the global FDL model among different SL groups with a proposed credibility weights prediction algorithm. Extensive experimental results demonstrate that compared with the baseline frameworks, the proposed IoV-SFDL framework achieves a 16.72\% reduction in edge-to-global communication overhead while improving about 5.02\% in model performance with the same training iterations.

\end{abstract}


\end{frontmatter}

\begin{figure*}[t]
\centering
\includegraphics[width=\textwidth]{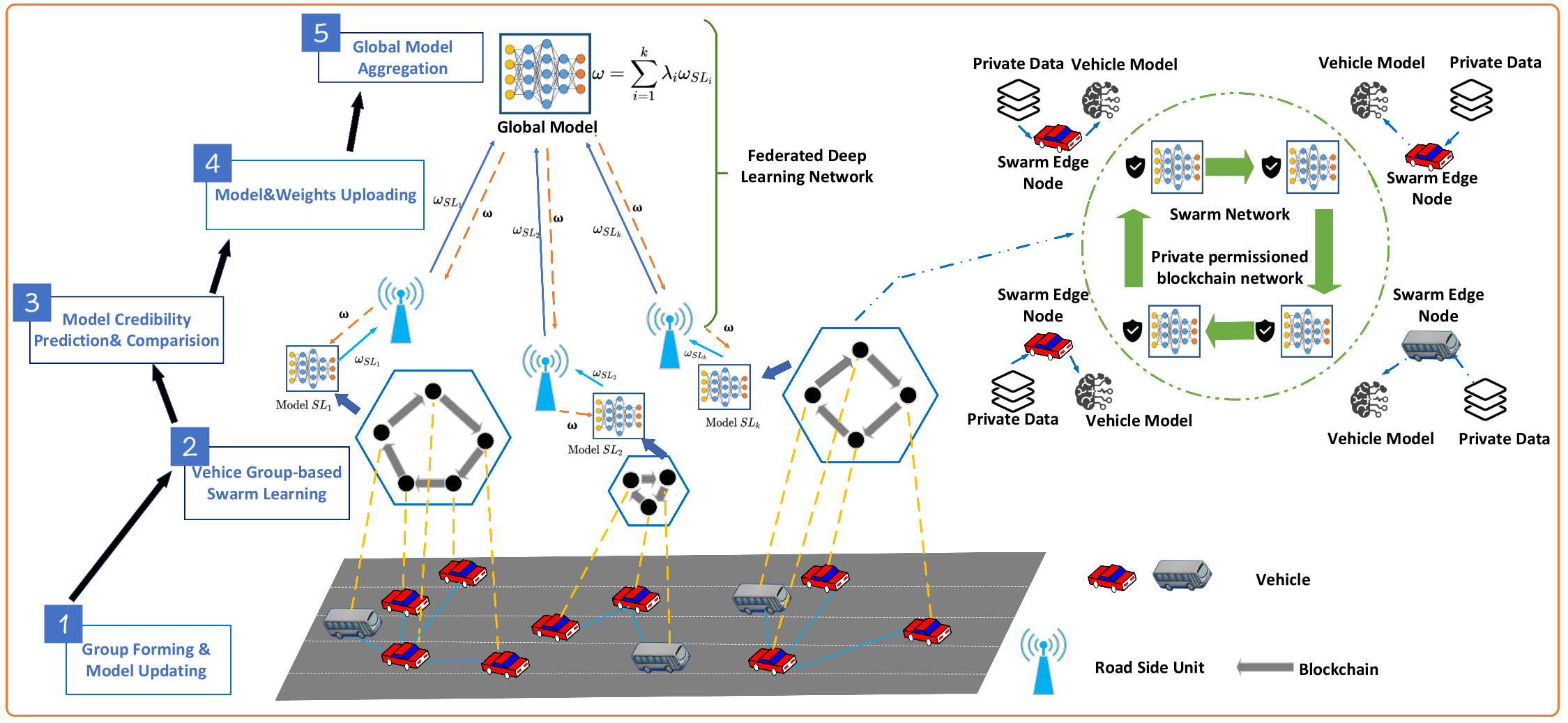}
\caption{An overview of the proposed IoV-SFDL framework. 
(1) Vehicles in the end-client layer form an SL group according to the relative distance and task type, update the latest global model in the group issued by Road Side Unit (RSU), collect surrounding information with onboard sensors, and train the model. 
(2) The vehicles in the SL group connect through blockchain and aggregate group model in the SL process.
(3) RSU nodes at the edge-server layer receive the model from SL groups and predict the weights according to group size and model credibility. 
(4) RSU nodes upload model parameters and weights to the central server. 
(5) The central server generates a global model by aggregating models from SL groups with different weights, and distributes the global model to the edge server for subsequent updates. After several iterations, the final model with good performance is obtained.
}
\titlespacing{\section}{0pt}{2.5ex plus 1ex minus .2ex}{1.3ex plus .2ex}

\label{fig:overview}
\end{figure*}

\section{Introduction}
With new computing and communication technologies coming up, the communication network's rapid development provides the possibility for advanced vehicular services and applications \cite{wang2021intelligent}, such as autonomous driving and content delivery. In this context, the Internet of Vehicles (IoV), as a combination of the Internet of Things (IoT) and Vehicular networks, has gained considerable attention from industry and academia. Some works in \cite{hamer2020fedboost,9448062,2021Reputation} have shown that compared with data centralized in traditional machine learning (ML), Federated Deep Learning (FDL) solves privacy concerns and reduces the cost of data transmission by distributing the training work to clients. Clients keep their data locally and send the model parameters to a central server for model aggregation\cite{li2020review}. In this way, FDL provides a parallel scheme to learn the global model cooperatively and protect data privacy \cite{hao2019towards}. 
% Federated Deep Learning (FDL) has been applied to IoV with its privacy protection and distributed features in the training process.

As for the FDL application in the IoV, it demands the efficiency of information exchanging about vehicle statuses and road conditions while protecting data privacy \cite{zhou2020evolutionary}. Scholars have made several communication efficiency improvements and achieved sound performance in IoV scenarios. For instance, \cite{2021FedVCP} proposed a Federated-Learning-Based Cooperative Positioning Scheme (FedVCP) to fully utilize the potential of collaborative edge computing while protecting data privacy in the IoV system. \cite{2021A} proposed a novel 
Federated Learning-based framework for Plate Recognition (FedLPR) to identify traffic signs and participants in low latency and consumption while keeping pace with preserving data privacy. \cite{zhang2019deep} proposed a two-timescale federated Deep Reinforcement Learning(DRL)-based algorithm to help obtain robust models in the IoV scenarios.

% In terms of communication security in IoV system,
\cite{fu2020privacy} integrated homomorphic cryptography into parking space searching and booking for communication security. \cite{wang2019privacy} designed a data collection and pre-processing scheme based on deep learning, reducing the amount of data uploaded to the cloud and effectively protecting the data privacy in the IoV system. Blockchain technology has been applied to IoV for data privacy in recent years as a newly emerging technology. For instance, 
\cite{shrestha2020new} created a local Blockchain for real-world event message exchange among vehicles within the boundary of the country, and \cite{lin2020blockchain} proposed a DRL and blockchain-empowered Spatial Crowd-sourcing System (DB-SCS). \cite{lu2020blockchain} developed a hybrid blockchain architecture that consists of the permission blockchain and the local Directed Acyclic Graph to relieve transmission load and address privacy concerns of data providers.

Based on blockchain technology and FDL, The authors in \cite{warnat2021swarm} introduced Swarm Learning (SL) and achieved outstanding performance in the medical field. The SL framework dispenses with a dedicated server, shares the parameters via the Swarm network, and builds the models independently on private data at the individual sites. However, all the above works don't consider vehicles' mobility, unreliable communication connection, and a highly dynamic driving environment. These factors will bring some new challenges for FDL and blockchain applications in the IoV system, including overcoming the model's ineffective and communication inefficiency caused by clients (such as malicious and redundant sharing parameters) and ensuring clients' privacy security from the third party in the process of data transmission \cite{CHEN2021}. 

This paper addresses these issues by proposing a Swarm-Federated Deep Learning framework in the IoV system (IoV-SFDL) with a credibility weights prediction algorithm. The contributions of this paper can be summarized as follows: 

(1) An IoV-SFDL framework is proposed to ensure data privacy with blockchain technology while minimizing the total cost of edge-to-global communication overhead, consisting of a central server, RSU-based edge servers, and vehicles. 

(2) A credibility weights prediction algorithm is proposed to improve the model convergence efficiency in the global model aggregation process.
	
(3) Various experiments in the IoV system are implemented to verify the performance of the IoV-SFDL framework and credibility weights prediction algorithm on the Next Generation Simulation (NGSIM) dataset.

The rest of this paper is organized as follows: The system architecture is present in Section 2. The proposed credibility weights prediction algorithm is detailed in Section 3.  Section 4 provides the NGSIM trajectory dataset, framework performance evaluation, and analysis. Section 5 concludes this paper.

\section{System Architecture}
% This section introduces the problem formation and the proposed IoV-SFDL framework. The main notations used throughout the paper are listed in Table \ref{tab:notation}, and the overview of the IoV-SFDL framework is shown in Figure \ref{fig:overview}.

The overview of the proposed IoV-SFDL framework is shown in Figure \ref{fig:overview}.  The problem formation, data transmission and framewrok detail of the proposed IoV-SFDL will be introduced in this section.

\begin{table}[t]
\centering  % 显示位置为中间
	\caption{SUMMARY OF MAIN NOTATIONS}  % 表格标题
	\label{table1}  % 用于索引表格的标签
\begin{tabular}{l|l}
\hline
\scriptsize $F(w)$ & \scriptsize The loss function of global model in cloud server    \\ \hline
\scriptsize $F_{SL}(w)$ & \scriptsize The loss function of the aggregated model in SL group   \\ \hline
\scriptsize$f_{i}(w)$ &\scriptsize The loss function of the model in the $i^{th}$ vehicle  \\ \hline
\scriptsize$W$& \scriptsize Aggregated model of central server    \\ \hline
\scriptsize$W_{SL}$& \scriptsize Aggregated model of SL group    \\ \hline
\scriptsize$L_m(i)$ & \scriptsize Local model parameters of the $i^{th}$ vehicle    \\ \hline
\scriptsize$R_m(i)$ & \scriptsize Merged model parameters of the $i^{th}$ vehicle  \\ \hline
\scriptsize$C_{b}(i)$ & \scriptsize The robustness value of the $i^{th}$ SL group  \\ \hline
\scriptsize$C_{e}(i)$ & \scriptsize The credibility value of the $i^{th}$ SL group  \\ \hline
\scriptsize$\lambda_{i}$ & \scriptsize The model weights of the $i^{th}$ RSU node \\ \hline
\scriptsize$D_{i}$ & \scriptsize 
The dataset obtained by onboard sensors of the $i^{th}$ vehicle \\ \hline
\scriptsize$e_{i}$ & \scriptsize The driving environment of the $i^{th}$ vehicle \\ \hline
\scriptsize$w^{i}$ & \scriptsize Onboard model of the $i^{th}$ vehicle \\ \hline
\scriptsize$\gamma^{i}$ & \scriptsize The model weights of the $i^{th}$ vehicle \\ \hline

\end{tabular}
\label{tab:notation}
\end{table}

\subsection{Problem Formation}
This paper assumes that a highway with $N$ vehicles facing different traffic conditions. The main notations used throughout the paper are listed in Table \ref{tab:notation}.
Because vehicular GPS, radars, cameras, and other onboard sensors can obtain real-time vehicle trajectory, we believe that each vehicle can be modeled as a triple $G=(V, P, E)$.  

(1) $V=\left\{v_{i} \mid i \in\{1, 2, \ldots\}\right\}$ represents all vehicles, where $v_{i}$ represents vehicle $i$ on the road. $N=|V|$ represents the number of vehicles on the road.

(2) $P=\left\{p_{i} \mid v_{i} \in V\right\}$ represents the private information of the vehicle, including the speed, acceleration, orientation, and position.

(3) $E=\left\{e_{i} \mid v_{i} \in V \right\}$ is a matrix, which is used to store the road environment information around the vehicle, that is, information of the vehicle from four directions in the visual distance.

The information $e_{i}$ of vehicle $v_{i}$ in road environment matrix is defined as:

\begin{equation}
\begin{aligned}
\mathrm{e}_{i} &=\left[\begin{array}{cc}
    I_{r} & I_{l} \\
    I_{p} & I_{f}
\end{array}\right]\\
\mathrm{I}\left(v_{i}, v_{j}\right) &= \begin{cases}0 & d\left(v_{i}, v_{j}\right)>R \\
1-\frac{\left(d\left(v_{i}, v_{j}\right)\right)^{2}}{R^{2}} & d\left(v_{i}, v_{j}\right) \leq R\end{cases},
\end{aligned}
\end{equation}
where $d\left(v_{i}, v_{j}\right)$ denotes the Euclidean Distance between $v_{i}$ and $v_{j}$, $R$ represents the visual range of vehicles.
% \textbf{Definition 2: }
% The private information $p_{i}$ acquired by the onboard sensors of the vehicle $v_i$ is a four tuple, which is defined as [$V_{s}(i), V_{a}(i), V_{o}(i), V_{p}(i)$], 
% % \begin{equation}
% % W_{i}=[V_{s}(i), V_{a}(i), V_{o}(i), V_{p}(i)]
% % \end {equation}
% where $V_{s}(i), V_{a}(i), V_{o}(i), V_{p}(i)$ represents the speed, acceleration, orientation and position of the vehicle $v_i$, respectively. 

\begin{figure*}[t]
\centering
\includegraphics[width=\textwidth]{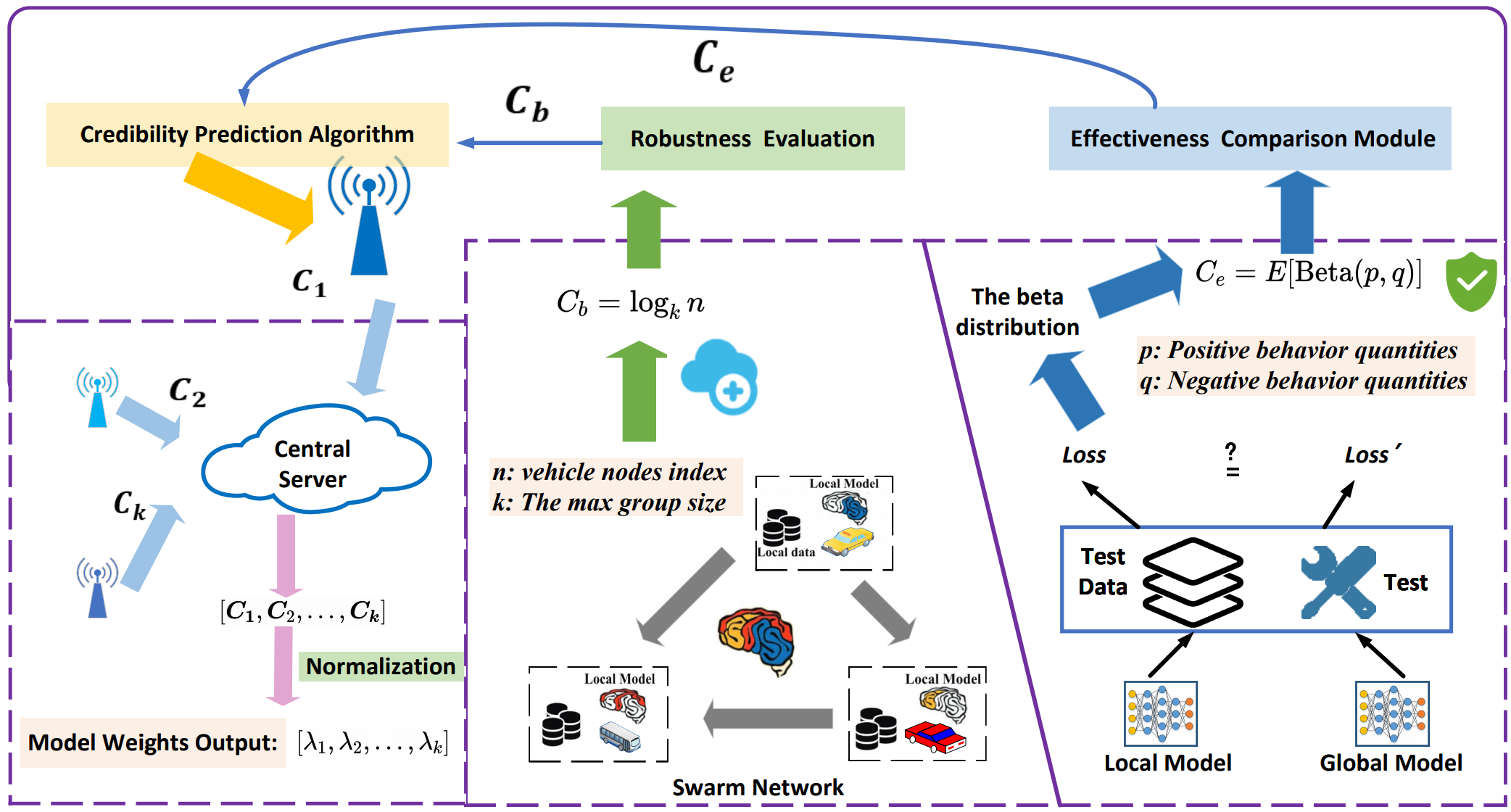}
\caption{The workflow in the proposed IoV-SFDL framework}
\label{fig:workflow}
\end{figure*}
% \begin{spacing}{1.0}

% \end{spacing}

Back propagation of the loss function is essential for ML training to adjust weights and parameters. The loss function in Deep Neural Network (DNN) models quantifies the modeling effect of the algorithm on the training data. The goal of training is to find the optimal parameters to minimize the loss function. The loss function in machine learning can be roughly divided into classification loss and regression loss. 
The general loss functions for classification include logarithmic loss, focus loss, Kullback-Leibler(KL) divergence, exponential loss, hinge loss, etc. 
This paper uses the Mean Square Error (MSE) as the loss function of the vehicle onboard model. The expression of the loss function $f_{i}(w)$ of the vehicle $v_i$ is as follows:
\begin{equation}
f_{i}(w)=\frac{\sum_{n=1}^{N}\left|{y}_{predict}^{n}-y_{observe}^{n}\right|^2}{N},
\end{equation}
where $y_{predict}^{n}$ and $y_{observe}^{n}$ represent the model prediction and observation of the sample batch $n$, and $N$ is the number of batches. 
Different functions are available for parameter merging as a configuration of the Swarm Learning, such as average, weighted average, minimum, maximum, or median functions. In this paper, the weighted average is defined as :
\begin{equation}
R_{m}(i)=\frac{\sum_{k=1}^{i}\left(\gamma^{k} \times L_{m}(k)\right)}{i \times \sum_{k=1}^{i} \gamma^{k}},
\end{equation}
% , and i is the number of vehicles participating in the SL group
in which $R_{m}(i)$ is merged parameters of the $i^{th}$ vehicle. $L_{m}(k)$ is model parameters from the $k^{th}$ vehicle, $\gamma$ is the model weight. In this way, the loss function of an SL group in this paper is defined as:
\begin{equation}
F_{SL}(w)=\frac{1}{\left|SL\right|} \sum_{i \in SL} [f_{i}(w)+f_{i-1}(w)],
\end{equation}
where $|SL|$ means the number of vehicles in an SL group.

Redundant or invalid experience data decrease the efficient implementation of the model aggregation process. According to the model’s effectiveness, the weights could be flexibly adjusted by the proposed credibility weights prediction algorithm. Making the model with better results have a higher weight in model aggregation helps the model converge faster and achieve higher accuracy in FDL. To describe the problem of model aggregation in IoV-SFDL, we introduce ${\lambda_i = \lambda_1, \lambda_2, \dots, \lambda_{|SL|}}$ as the weights in global model aggregation. $|SL|$ represents the number of SL groups on the road.
The objective function of global learning optimization is defined as:
\begin{equation}
F(w)=\frac{1}{\left|SL\right|} \sum_{i \in SL} \lambda_{i}F_{{SL}_i}(w).
\end{equation}

As mentioned earlier, we use the minimization of the loss of the global model for the current training task as the objective function:
\begin{equation}
min\,F(w)=min\,[\frac{1}{\left|SL\right|} \sum_{i \in SL} \lambda_{i}F_{{SL}_i}(w)].
\end{equation}

\subsection{Swarm-Federated Deep Learning Framework}
The IoV-SFDL framework in the IoV system contains three layers: A central server in the cloud layer, RSU-based edge nodes in the edge layer, Vehicle clients in the end layer. The three-layer structure constitutes the whole IoV-SFDL architecture, and the operation is divided into three steps as a whole. First of all, vehicles in the end layer collect trajectory information and train a time-sensitive local model. According to the SL process, vehicles with the same training task will form a blockchain network, with local model parameters transferred and aggregated. Then, the RSUs receive the model parameters from the SL group and predict model weights according to the credibility weights prediction algorithm. At last, the parameters and weights are transmitted to the central server for a global model aggregation. 

The model's aggregation process can be divided into local SL and global FDL processes. Each vehicle aggregates the model passed from the previous vehicle with its onboard model in the local swarm learning process. In this process, the distributed feature of SL is combined with the point-to-point network based on blockchain. Compared with the general operation of federated deep learning, it provides a higher level of data security, central coordinator elimination, and machine learning model attacks protection. According to the proposed credibility weights prediction algorithm, each RSU predicts the model's weight in the global FDL process and sends the model parameters to the central server. In this process, the global FDL realizes the global model aggregation through a central server, which aggregates the value of local model parameters according to their weights to generate a global model. Compared with the general operation of FDL, the IoV-SFDL provides less edge-to-global communication overhead and a more effective training process. 

Above all, the data transmission scheme in the framework consists of five steps: (1) Group forming and model updating; (2) Vehicle group-based Swarm Learning; (3) Credible weights prediction and comparison; (4) Model parameters and weights uploading; (5) Global Model Aggregation. After several iterations of these steps, the final model with good performance is obtained.

% Above all, the data transmission scheme in the framework consists of five steps:
% Group forming and model updating
% Vehicle group-based Swarm Learning
% Credible weights prediction and comparison
% Model parameters and weights uploading
% Global Model Aggregation
% The data transmission forms a closed-loop; after several iterations, the final model with good performance is obtained.

% 1) Group forming and model updating: vehicles collect vehicle trajectory and traffic environment information to train a time-sensitive model. 

% 2) Vehicle group-based SL: After completing the local model training. The onboard vehicle model aggregates with models transmitted from other vehicles with the same training task through blockchain. 

% 3) Credible weights prediction and comparison: Model parameters sent from the SL group are evaluated for their effectiveness and robustness at the RSU. With the help of historical experience, the weight of the model is predicted and updated. 

% 4) Model and weights uploading: The RSU nodes upload model parameters and weights to the central server for global model aggregation.

% 5) Global Model Aggregation.: The central server aggregates all the model parameters sent by RSU according to the model weights.

\section{Credibility Weights Prediction Scheme}
The key to guaranteeing model security and training efficiency in the IoV system is detecting ineffective models and scheduling less of them in each communication round. This work proposes a credibility weights prediction algorithm in the IoV-SFDL framework to prevent unreliable updates and minus noise from models with poor effectiveness. With the credibility weights prediction algorithm, the RSU node will evaluate the model’s effectiveness and robustness of the received model and global model. The model’s credibility will be predicted as the weight for the global model aggregation process in each communication round. The total workflow is shown in Figure \ref{fig:workflow}. The algorithm can help the central server in FDL distinguish effective models from ineffective ones, thereby preventing malicious or incorrect updates of the global aggregate process in FDL.

\subsection{Credibility Weights Prediction Algorithm}
The Credible value $C_{i}$ of model $i$ in this paper can be divided as $C_{b}$ and $C_{e}$, where $C_{b}$ and $C_{e}$ represent the robustness value and effectiveness value of model from an SL group.
To represent the node’s robustness, we consider the logarithms’ formulation, a single increasing function.
In ML, we believe that the more datasets in the training process, the better robustness of the model will be. Therefore, in global model aggregation, the model trained on more datasets can be given higher weight. Thus, the robustness value of the model with n vehicles in SL is given by the following formula:
\begin{equation}
C_{b}=\log _{k} n.
\end{equation}
In this formula, we believe that $n$ is the number of vehicles in the current SL group, $k$ is the max group size among all SL groups. The specific experimental parameters will be classified in the next section.

To represent the model's effectiveness, we use observation $B$ made by the RSU node to illustrate the effective prediction $P(e_i)$ of a model. 
The effectiveness of the model, $P\left(e_i \mid B\right)$ is calculated and predicted by the RSU node based on the observation, which can be derived as:
\begin{equation}
P\left(e_i \mid B\right)=\frac{P\left(e_i\right) P\left(B \mid e_{i}\right)}{\sum_{j=1}^{n} P\left(e_{j}\right) P\left(B \mid e_{j}\right)},
\end{equation}
To facilitate the expression and update of effectiveness value, we use beta distribution to represent aggregate weights from the effectiveness of an RSU node, defined as:
\begin{equation}
\begin{aligned}
f(x ; p, q)=\frac{x^{p-1}(1-x)^{q-1}}{\int_{0}^{1} u^{p-1}(1-u)^{p-1} d u}\\\\
=\frac{\Gamma(p+q)}{\Gamma(p) \Gamma(q)} x^{p-1}(1-x)^{q-1}\\\\
=\frac{1}{B(p, q)} x^{p-1}(1-x)^{q-1}
\end{aligned},
\end{equation}
where $ p>0 \text { and } q>0 $ and p and q respectively represent the number of times that the receiving model is better than the global model and the number of times that the global model is better than the receiving model, the $\Gamma(x)$ is the Gamma function, which could be written as:

\begin{equation}
\Gamma(x)=\int_{0}^{+\infty} t^{x-1} e^{-t} \mathrm{~d} t(x>0),
\end{equation}

For example, we assume the $a_{i}$ and $b_{i}$ represent the positive and negative behaviors currently, while p and q represent the initial positive and negative behaviors.
To update the credibility, it is equivalent to updating the two parameters p and q, as:
\begin{equation}
\left\{\begin{array}{l}
p^{\text {new }}=a_{i}+p \\
q^{\text {new }}=b_{i}+q
\end{array}\right.,
\end{equation}

When a node is initialized without prior knowledge, the credibility of a node can be expressed as a uniform distribution on $(0,1)$, defined as:

\begin{equation}
P(x)=\operatorname{uni}(0,1)=\operatorname{Beta}(1,1).
\end{equation}
By comparing the effectiveness of the receiving model and the global model previous time, the RSU can evaluate the effectiveness of the receiving model, that is, convert into positive behavior and negative behavior. The evaluation process of the models is detailed in the next section. Suppose that in $n$ rounds of communication, the SL group has sent the model to RSU $(p+q)$ times. By comparing the effectiveness of the models, these interactions are described as $p$ times positive behavior and $q$ times negative behavior. With this information, the RSU can predict the model's effectiveness from SL group in the next event. The validity probability of the model is defined as:
\begin{equation}
C_{e}=E[Beta(p,q)]=\frac{p}{p+q}.
\end{equation}

Central server can normalize all the weights and model parameters sent by the RSUs, and the weights of model from $i^{th}$ RSU is defined as:
\begin{equation}
\lambda_{i}=\frac{C_{i}}{\sum_{j=1}^{n} C_{j}}.
\end{equation}

\subsection{Effectiveness Comparison Module}

In the model aggregation process of FDL, the credibility comparison module at RSU uses the test dataset to evaluate and compare the model received from SL group and global model download from central server. Based on the value of the loss function derived from the test dataset, we define the model's credibility for comparing the effectiveness of different models in model aggregation.

\begin{equation}
    e_{k}^{t}=\frac{f(w^{t}, {x}_{i j}, y_{i j}) - f(W_{k}^{t}, {x}_{i j}, y_{i j})}{f({w}^{t}, {x}_{i j}, y_{i j})},
\end{equation}
where $\left({x}_{i j}, y_{i j}\right)$ is a data sampled from the current RSU node’s test dataset. We use the difference between the loss function value of the SL model and global model to evaluate the quality of the received SL model parameters. 

For a tagged RSU node, its contribution to the global model aggregation can be considered positive or negative. Precisely, if $e_{k}^{t} > 0$, we assume that the effectiveness comparison module will characterize RSU node $k$'s transmission as positive behavior and vice versa. Moreover, we can also use $e_{k}^{t}$ to represent the effect gap between the local model and the receiving model imposed by the RSU. 

\subsection{Vehicle Group-based Swarm Learning}
In the process of local model aggregation, vehicle group-based swarm learning is considered. RSUs, as the participants in the local model aggregation process, are evenly distributed on both sides of the road. The SL framework cooperatively optimizes the model inside the vehicle through continuous iteration of the following three steps: 

1) Each vehicle carries out the model training task by collecting the privacy information of local vehicles, such as geographical location, speed, direction angle, etc. 

2) Vehicles with the same subtask and close to each other can form groups and aggregate models through blockchain in the SL process. 

3) After model aggregation, the vehicle sends the model parameters to the RSU for the further procession.

Generally, the number of vehicles in different SL groups is different due to the rapid mobility of vehicles on the road and the uncertainty of driving purpose. By evaluating the credibility of the SL groups and their internal LSTM model effectiveness, the best weight in the global FDL aggregation process could be predicted. The whole training process of vehicle group-based SL is shown in Algorithm \ref{alg:swarm_learning}.

\begin{algorithm}[htb]
    \caption{Vehicle Group-based Swarm Learning. The $K$ vehicles are indexed by $k$; $D_{k}$ is local dataset of vehicle $k$, $\eta$ is the learning rate, $B$ is the local mini-batch size, $w_{t}^{k}$ is the onboard model of the vehicle, and $W_{SL}$ represented the aggregated model of SL group.} 
    \label{alg:swarm_learning}
    \LinesNumbered
    \hspace*{0.02in} {\bf{Input:} $w_{t}^{k}$}\\
    \hspace*{0.02in} {\bf{Output:} $W_{SL}$}
    Initialize  $w_{0}^{k}$ \;
    \For{each round t = 1, 2, \dots}{
        Update model ${w}_{t}^{k}$ from RSU\;
        $b \leftarrow$ spilt $D_{k}$ into batches of size $\mathcal{B}$\;
        \For{each vehicle k = 1, 2, \dots } { 
            $w_{t+1}^{k} \leftarrow  w_{t}^{k}-\eta \nabla \ell(w_{t}^{k}; b)$\;
            $w_{t+1}^{k} \leftarrow \frac{w_{t+1}^{k}+ w_{t+1}^{k-1}}{2}$\;
        }
    }
    $W_{SL} = w_{t+1}^{k}$\;
    \Return $W_{SL}$\;
\end{algorithm}

\subsection{Global Federated Deep Learning}
In the global model aggregation process, an FDL framework with credibility is considered. As shown in Figure \ref{fig:overview}, the FDL enables the local RSUs and central server to collaboratively optimize the global model by continuously iterating the following two steps: 

1) Each RSU receives the model parameters, predicts the weight by the credibility weights prediction algorithm, and sends the parameters to the central server in the up-link.

2) The central server collects the model parameters from RSU, aggregates the global model according to their weights predicted from RSU, and broadcasts the new model to all the RSUs in the down-link. Along the training process, each iteration is generally referred to as a communication round. 

After sufficient communication rounds involving parameters exchanges and training, we can obtain the best global FDL model without sacrificing the vehicle’s privacy. The whole training process of FDL is shown in Algorithm \ref{alg:global_federated_Deep_Learning}.

\begin{algorithm}[htb]
    \caption{Global Federated Deep Learning. The K RSUs are indexed by k; $w^{k}_t$ is local model parameters at time t, $\lambda_{t}^{k}$ is model weights of $w^{k}_t$, $W_t$ is the global model parameters} %算法的名字
    \label{alg:global_federated_Deep_Learning}
        \hspace*{0.02in} {\bf{Input:} $w^{k}_t$, $\lambda^{k}_t$} \\
        \hspace*{0.02in} {\bf{Output:} $W$}
        Initialize $w^{k}_0$\;
        \For{each round t = 1, 2, \dots}{
            \For{each RSU k}{
                $\lambda_{t+1}^{k} \leftarrow WeightsPrediction({w}^{k}_{t})$\;
            }
            $W \leftarrow \sum_{k=1}^{K} \frac{1}{K} \lambda_{t+1}^{k} w_{t+1}^{k}$\;
        }
        \Return $W$
\end{algorithm}

\begin{figure*}[t]
\centering
\includegraphics[width=\textwidth]{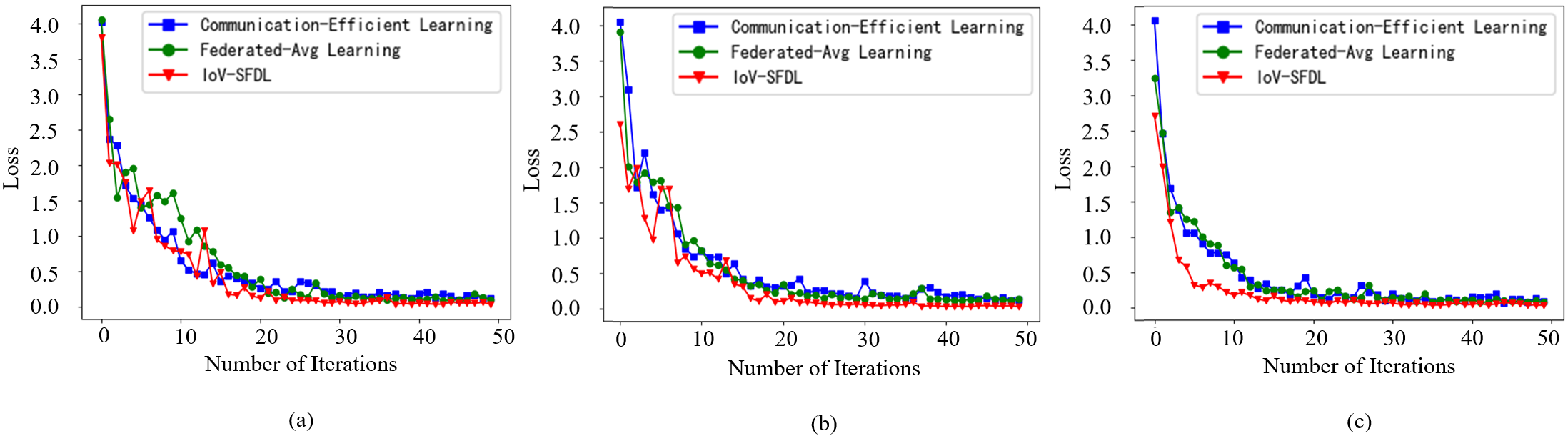}
\caption{Loss comparison of the IoV-SFDL, the Federated-Avg Learning in \cite{li2020federated}, and the Communication-Efficient Learning in \cite{mcmahan2017communication} on different traffic densities. (a)High-density traffic: 10 vehicles in one group and 6 vehicles in the other group. (b)Medium-density traffic: 6 vehicles in one group and 4 vehicles in the other group. (c)Low-density traffic: 2 vehicles in one group and 3 vehicles in the other group.}
\label{fig:loss_compare}
\end{figure*}

\section{Performance Evaluation}

In this section, we introduce the simulation setups and then compare the performance of the proposed IoV-SFDL with the other two baseline frameworks.  All source code about the framework, dataset used, and proposed algorithm in this study are provided on GitHub (https://github.com/CoderTylor/IoVSFDL-Swarm-Federated-Deep-Learning-).

\subsection{Experiment Setup}
The hardware is 32GB RAM, i7-10700 CPU 2.90GHz. 
The experimental platform is built based on the PyTorch 1.9 framework, a flexible deep learning tool widely used in academics.  The software adopts Python 3.7.0 in the Ubuntu18.04 system.  
Inspired by the performance of Social LSTM \cite{alahi2016social}, an LSTM-centered model is trained as an onboard vehicle training task to predict the vehicle’s trajectory in 5s. The complete model parameters are listed in Table \ref{tab:LSTM}. For all LSTM learning models, we used the greedy exploration with linearly annealed from 1.0 to 0.1 in the whole training process. Adam optimizer is used to learn all neural network parameters with a learning rate of $10^{-4}$. 

\begin{table}[b]
\caption{LSTM Model Parameters}
\centering
\begin{tabular}{cc}
% \caption{Neural Network parameters}
\hline
\footnotesize CELL                   & \footnotesize (Input, Output) \\ \hline
\footnotesize LSTM CELL              & \footnotesize (64,32)  \\\hline
\footnotesize Input Embedding Layer  & \footnotesize (In Feature=9, Out Feature=32) \\\hline
\footnotesize Social Tensor Conv1    & \footnotesize (32,16, Kernel size=(5,3), stride(2,1)) \\\hline
\footnotesize Social Tensor Conv2    & \footnotesize (16,8, Kernel size=(5,3), stride(2,1))  \\\hline
\footnotesize Social Tensor Embedded & \footnotesize (In Feature=32, Out Feature=32)        \\\hline
\footnotesize Output Layer           & \footnotesize (In Feature=32, Out Feature=5)          \\\hline
\footnotesize Dropout          & \footnotesize Dropout(p=0, inplace=False)         \\\hline
\footnotesize Activation Function          & \footnotesize ReLU         \\\hline
\footnotesize Time Length          & \footnotesize 10        \\\hline
\footnotesize Episode          & \footnotesize 50        \\\hline
\footnotesize Learning rate          & \footnotesize $10^{-4}$          \\\hline
\footnotesize Optimizer          & \footnotesize Adam          \\\hline
\end{tabular}
\label{tab:LSTM}
\end{table}

The simulation defines three traffic flow densities: high-traffic flow density, medium-traffic flow density, and low-traffic flow density. The number of total vehicles ranges from 5 to 16, and the size of the SL group ranges from 2 to 10. The NGSIM trajectory dataset is used to verify the effectiveness of the IoV-SFDL framework. The dataset is collected for the next-generation traffic simulation by Federal Highway Administration. It use videos to capture real-world traffic information, including vehicle velocity, position, acceleration, lane, etc. As high-resolution real-world vehicle trajectory data, NGSIM is widely used to explore the characteristics of the trajectory-prediction process and calibrate and validate the trajectory-prediction models. Detailed descriptions of the dataset can be found in \cite{coifman2017critical}.

To the best of our knowledge, there is no research investigating the SL application under the IoV system, so two other baseline FDL frameworks are added in the simulations for  better gauging the performance of the proposed IoV-SFDL framework, which are:

(1) Federated-Average Learning (Federated-Avg): Based on the Federated Average deep learning algorithm proposed in \cite{li2020federated}, all vehicles on the road participating in FDL. After training the model locally, the average aggregation of model parameters is carried out in the central server. 

(2) Communication-Efficient Federated Learning (Communication-Efficient): Based on the Communication-Efficient federated deep learning algorithm proposed in \cite{mcmahan2017communication}, we add the LSTM network for trajectory prediction in their code and set the parameters $Frac$ as 0.8 for client selection.

\subsection{Indicator}
Three indicators are chosen to comprehensively evaluate the performance of the proposed IoV-SFDL framework, such as model Loss, trajectory Prediction Accuracy, and Prediction Error. 
The descent speed of the Loss function is the first indicator to evaluate the converging rate. The Loss of the LSTM model is defined as:
\begin{equation}
{\rm Loss}=\small \frac{\sum_{i=1}^{N}\left|(x_{p r e d i c t}^{i},y_{p r e d i c t}^{i})-(x_{observe}^{i},y_{observe}^{i})\right|^{2}}{N} 
\end{equation}
where $(x_{predict},y_{predict})$ is the vehicle trajectory predicted by LSTM, and $(x_{observe},y_{observe})$  is the actual trajectory of the vehicle in the dataset.

Then, the Prediction Error is used to evaluate the average Euclidean distance between predict trajectory and actual trajectory in the NGSIM dataset under different traffic densities in meters. The Prediction Error is defined as:
\begin{equation}
\begin{aligned} 
\rm \tiny Pre& \rm diction \ Error=\\\\&\small \frac{\sum_{i=1}^{N}\sqrt{(x_{predict}^{i}-x_{observe}^{i})^{2}+(y_{predict}^{i}-y_{observe}^{i})^{2}}}{N}.
\end{aligned}
\end{equation}

At last, the Predict Accuracy is evaluated to estimate the model performance.
For the trajectory after 5 seconds, we compare the distance difference of each path point as the basis for whether the prediction is true or false. To be precise, we define positive prediction $P_{T}$, negative prediction $P_{F}$ and prediction accuracy as:
\begin{equation}
\text {P}= \begin{cases}\text { $P_{F}$ } & {\rm Prediction \ Error}>10 m \\ \text { $P_{T}$ } & {\rm Prediction \ Error}\leq 10 m\end{cases}.
\end{equation}
Based on equation (18), $N(P=P_{T})$ and $N(P=P_{F})$ is defined as the number of positive and negative prediction, and the Prediction Accuracy is defined as:

\begin{equation}
{\rm Prediction \ Accuracy}= \small \frac{N(P=P_{T})}{N(P=P_{T})+ N(P=P_{F})}.
\end{equation}

% \begin{figure*}[t]
% \centering
% \includegraphics[width=\textwidth]{Loss.png}
% \caption{Loss comparison of the IoV-SFDL, the Federated-Avg Learning in \cite{li2020federated}, and the Communication-Efficient Learning in \cite{mcmahan2017communication} on different traffic densities. (a)High-density traffic: 10 vehicles in one group and 6 vehicles in the other group. (b)Medium-density traffic: 6 vehicles in one group and 4 vehicles in the other group. (c)Low-density traffic: 2 vehicles in one group and 3 vehicles in the other group.}
% \label{fig:loss_compare}
% \end{figure*}

% \subsection{Performance of the IoV-SFDL Framework}

\begin{figure*}[]
\centering
\includegraphics[width=\textwidth]{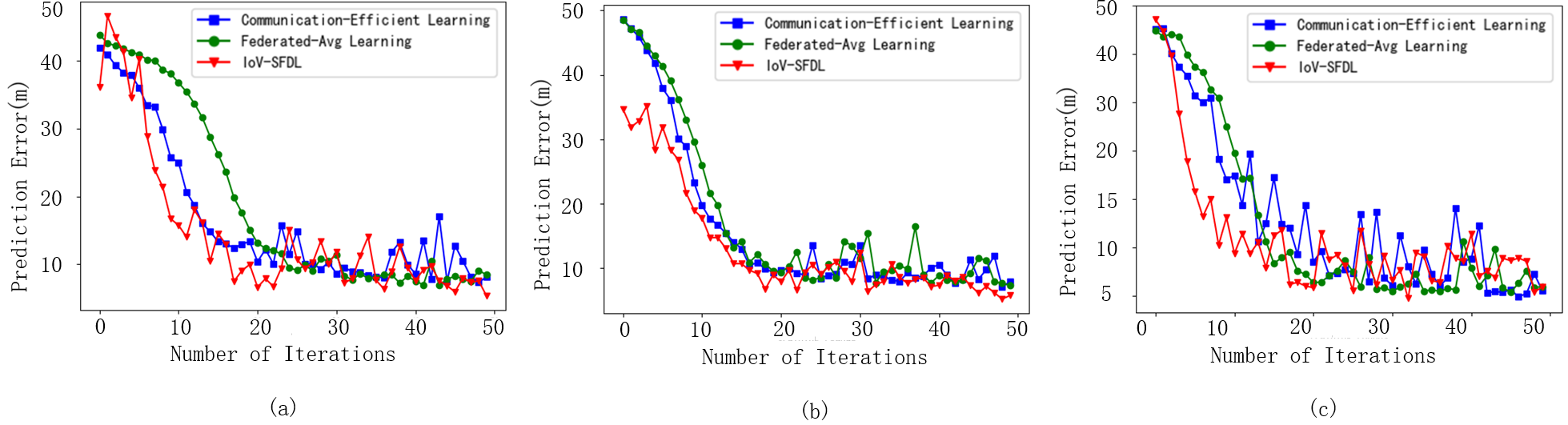}
\caption{Prediction Error comparison of the IoV-SFDL, the Federated-Avg Learning in \cite{li2020federated}, and the Communication-Efficient Learning in \cite{mcmahan2017communication} on different traffic densities. (a)High-density traffic: 10 vehicles in one group and 6 vehicles in the other group. (b)Medium-density traffic: 6 vehicles in one group and 4 vehicles in the other group. (c)Low-density traffic: 2 vehicles in one group and 3 vehicles in the other group.}
\label{fig:pred_err_compare}
\end{figure*}

\begin{figure*}[]
\centering
\includegraphics[width=\textwidth]{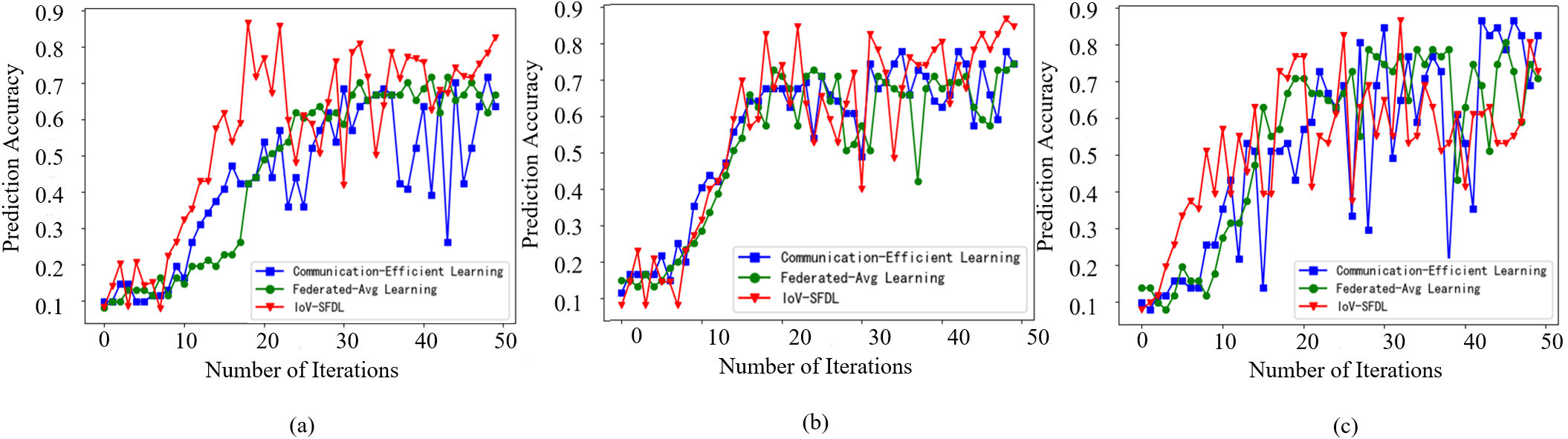}
\caption{Prediction Accuracy Comparison of the IoV-SFDL, the Federated-Avg Learning in \cite{li2020federated}, and the Communication-Efficient Learning in \cite{mcmahan2017communication} on different traffic densities. (a)High-density traffic: 10 vehicles in one group and 6 vehicles in the other group. (b)Medium-density traffic: 6 vehicles in one group and 4 vehicles in the other group. (c)Low-density traffic: 2 vehicles in one group and 3 vehicles in the other group. }
\label{fig:pred_acc_compare}
\end{figure*}

% \begin{figure}[t]
% \centering
% \includegraphics[width=7cm]{Communication links.png}
% \caption{Communication Links of Different Frameworks}
% \label{fig:comm_link}
% \end{figure}

\subsection{Performance of the IoV-SFDL Framework}

In Figure \ref{fig:loss_compare}, we can see that the loss function change is almost the same in the first ten episodes among IoV-SFDL and the other two benchmark frameworks. As shown in Figure \ref{fig:loss_compare}(b) and \ref{fig:loss_compare}(c), the IoV-SFDL performs better on the same computing cost than the others. Although Figure \ref{fig:loss_compare}(a) shows that the change speed of the loss function of the three frameworks is almost the same under the condition of high-density traffic flow, the loss function of IoV-SFDL is lower than the other two benchmark frames after 20 iterations. This is because, with the increase of traffic flow, the proposed credibility weight prediction algorithm can help the cloud server identify the redundant model update, And give this update a minor weight to ensure the convergence speed of the model.

Figure \ref{fig:pred_err_compare} shows the Prediction Error between predict trajectory and actual trajectory under different traffic densities in meters. With the same experimental parameters settings of the three frameworks, the Prediction Error decrease with the iteration number increase, indicating that the LSTM learning agent has learned to predict trajectory more precisely. As shown in Figure \ref{fig:pred_err_compare}(a), \ref{fig:pred_err_compare}(b), and \ref{fig:pred_err_compare}(c), the more vehicle clients participating in the FDL, the redundant data will affect the model prediction error of the two baseline frameworks. Federated-Avg Learning is the most vulnerable framework facing redundant training data. In contrast, the IoV-SFDL can converge with less Prediction Error quickly under all traffic densities given. 

Figure \ref{fig:pred_acc_compare} shows the model’s Prediction Accuracy comparison for trajectory under the IoV-SFDL framework with the other two baseline frameworks. We use equation 18 to classify the positive and negative values of the predicted points so that the LSTM becomes a binary network. Figure \ref{fig:pred_acc_compare}(c) illustrates that the Prediction Accuracy difference between the IoV-SFDL framework and baselines in Prediction Accuracy is not evident under low-density conditions. However, as shown in Figure \ref{fig:pred_acc_compare}(a) and \ref{fig:pred_acc_compare}(b), with more vehicles clients participating in global model aggregation, the IoV-SFDL can achieve higher model accuracy with the same computing cost.

\begin{figure}[t]
\centering
\includegraphics[width=7cm]{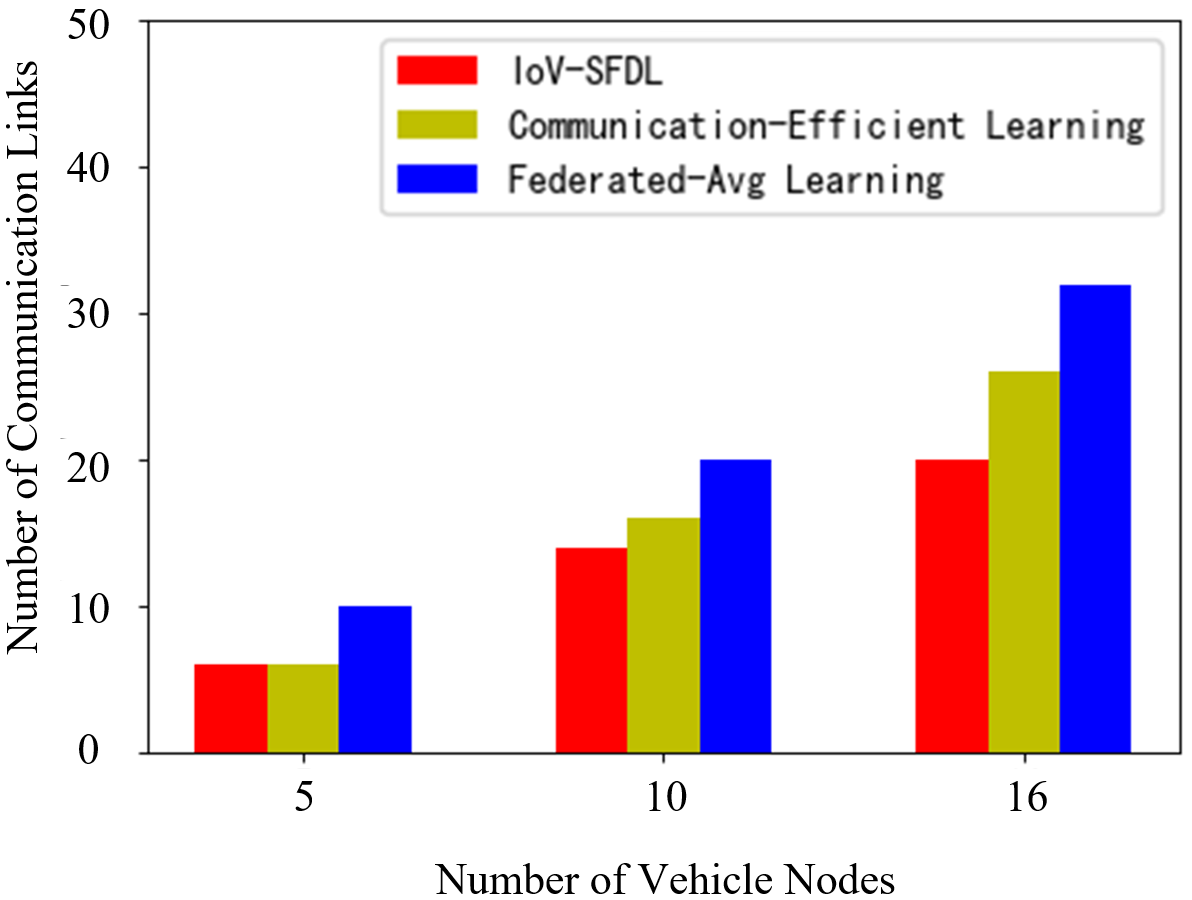}
\caption{Communication Links of Different Frameworks}
\label{fig:comm_link}
\end{figure}
Figure \ref{fig:comm_link} shows the communication links establishment difference between IoV-SFDL and baseline frameworks. Unlike Fed-Avg Learning and Communication-Efficient Learning, each client participating in global model aggregation has to establish a two-way link with the central server. The IoV-SFDL only needs to establish a blockchain connection between clients in the SL group. Each group merely requires one two-way communication link with the central server to realize global model aggregation. Under the conditions of 5, 10, 16 vehicles (Low-density, Medium-density, and High-density) on the road, compared with the other two baseline frameworks, the IoV-SFDL needs the least communication links to complete the global model aggregation. 

Overall, the proposed IoV-SFDL framework is superior to other baseline frameworks in edge-to-global communication overhead and convergence times. 

\section{Conclusion}
This paper proposed an IoV-SFDL framework for protecting data privacy by integrating the parallel SL into the FDL process.  
With the help of the SL process, the reliability of shared data among FDL is guaranteed by blockchain verification. 
Moreover, to achieve the high-level performance of the proposed framework in the IoV system, a credibility weights prediction algorithm was designed to speed up the convergence of model training in the IoV scenery. Simulation results showed that the IoV-SFDL framework has less edge-to-global communication overhead and higher computing efficiency than other state-of-art FDL frameworks in the IoV system. 
Our future work will expand the scale of the experiment and show how our framework protects vehicle privacy data in the case of attacks.

\bibliographystyle{elsarticle-num}
\bibliography{egbib}
~~~\\
~~~\\

\end{document}